\documentclass[12pt]{amsart}

\usepackage{amssymb,verbatim,graphicx}
\usepackage[mathscr]{eucal}
\usepackage{enumerate}
\usepackage{xspace}
\usepackage[thinlines]{easymat}
\usepackage{tabularx}

\newenvironment{evidence}{\begin{trivlist} \item[] {\bf Evidence:} \ }{\hfill
\qed \end{trivlist} }

\newtheorem{theorem}{Theorem}[section]

\newtheorem{fact}[theorem]{Fact}

\theoremstyle{definition}

\newtheorem{remark}[theorem]{Remark}

\numberwithin{equation}{section}

\def\&{\wedge}

\begin{document}

\title{Compactness statistics for spanning tree recombination}

\author{Jeanne N. Clelland}
\address{Department of Mathematics, 395 UCB, University of
Colorado, Boulder, CO 80309-0395}
\email{Jeanne.Clelland@colorado.edu}

\author{Nicholas Bossenbroek}
\email{Nicholas.Bossenbroek@colorado.edu}

\author{Thomas Heckmaster}
\email{Thomas.Heckmaster@colorado.edu}

\author{Adam Nelson}
\email{Adam.M.Nelson@colorado.edu}

\author{Peter Rock}
\email{Peter.r.rock2@gmail.com}

\author{Jade VanAusdall}
\email{jade.vanausdall@gmail.com}

\subjclass[2010]{60J10, 05C70, 91F10}
\keywords{Redistricting, gerrymandering, Markov Chain Monte Carlo (MCMC), ReCom sampling method}

\thanks{This paper is the result of an REU project sponsored by the Department of Mathematics at the University of Colorado Boulder during Summer 2019.
}
\thanks{The first author was supported in part by a Collaboration Grant for Mathematicians from the Simons Foundation.}

\begin{abstract}
Ensemble analysis has become an important tool for analyzing and quantifying gerrymandering; the main idea is to generate a large, random sample of districting plans (an ``ensemble") to which any proposed plan may be compared.  If a proposed plan is an extreme outlier compared to the ensemble with regard to various redistricting criteria, this may indicate that the plan was deliberately engineered to produce a specific outcome.

A variety of methods have been used to construct ensembles of plans, and a fundamental question that arises is: How likely are various types of plans to appear in an ensemble constructed by a particular method, and what types of plans are most likely to appear?  More precisely, given a method for constructing plans, is it possible to identify a probability distribution on the space of plans that describes the probability of constructing any particular plan by that method?

Recently, Markov Chain Monte Carlo (MCMC) methods have become a predominant tool for constructing ensembles of plans.  In this paper, we focus on the MCMC method known as ``ReCom," which was introduced in 2018 by the MGGG Redistricting Lab.  This method tends to produce plans with relatively compact districts compared to some other methods, and we sought to understand this phenomenon in greater detail.  To this end, we adopted a discrete analog of district perimeter called ``cut edges" as a quantitative measure for district compactness; this measure was
proposed by Duchin and Tenner, and it avoids some of the difficulties associated with more traditional compactness measures based on geographic perimeter, such as the Polsby-Popper score.

In order to model the basic ReCom step, we constructed large ensembles of plans consisting of two districts for two grid graphs and for the precinct graph of Boulder County, CO. We found that the probability of sampling any particular plan---which is known to be approximately proportional to the product of the numbers of spanning trees for each of the two districts---is also, to a high degree of accuracy, proportional to an exponentially decaying function of the number of cut edges in the plan.  This represents the first quantitative evidence of this specific relationship between spanning trees and cut edges, and it is an important step towards understanding compactness properties for districting plans produced by the ReCom method.

\end{abstract}

\maketitle

\section{Introduction}

Partisan gerrymandering has emerged as a fundamental issue in the periodic redrawing of legislative districts at all levels, from U.S. Congressional districts down to municipal districts for city councils, school boards, etc. The availability of sophisticated data and software tools has enabled the drawing of district lines so as to advantage one political party much more precisely and effectively in recent years than was possible previously, and this in turn has led to a flood of litigation challenging district plans on the basis of excessive partisan gerrymandering.  Unlike racial gerrymandering, which has historically been limited by the Voting Rights Act of 1965, partisan gerrymandering has largely been unchecked by the courts until fairly recently, primarily due to the difficulty of identifying a quantifiable standard for measuring it.  The National Conference of State Legislatures report \cite{NCSLSupremeCourt} provides an excellent summary of the case law regarding redistricting as of Spring 2019.

In June 2019, the U.S. Supreme Court declared in {\em Rucho v.~Common Cause} \cite{Rucho} that claims of partisan gerrymandering will no longer be considered justiciable in federal courts.  However, some claims have been brought successfully in state Supreme Courts; for instance, in January 2018, the Pennsylvania Supreme Court ruled \cite{LWV} that the state's 2011 districting map, widely considered a Republican gerrymander, violated the Pennsylvania state constitution.   After the {\em Rucho} ruling, later in 2019 the Wake County Superior Court unanimously struck down North Carolina's legislative maps as unconstitutional \cite{CommonCause}, ruling that the maps violated the state constitution's guarantees of free elections, equal protection, freedom of speech and freedom assembly.  With the next round of redistricting coming in 2021, many more such cases are certain to be litigated in state courts in the coming years.

A fundamental question that must be addressed is: How can partisan gerrymandering be definitively identified and measured quantitatively?  Courts have been largely unconvinced by simple, one-size-fits-all measures such as partisan symmetry (introduced in 1987 in \cite{KB87}) and efficiency gap (introduced in 2014 in \cite{SM14}) that fail to take into account variations in political geography that can vary widely between states.  Gerrymandering is simply too complicated a phenomenon to be adequately described by any single measure. 

One strategy that has recently gained traction is to compare a particular districting plan to a large collection of randomly generated plans.
The idea of using computers to generate random districting plans dates back to the 1960's; more recently, in \cite{CR13} Jowei Chen and Jonathan Rodden introduced the idea of building a collection of random plans and using them to compare a proposed plan to a variety of random alternatives.  
This approach was substantially refined in expert work of Jonathan Mattingly  to a statistical outlier approach via Markov Chain Monte Carlo methods for generating large ensembles of plans \cite{NC-cong-report}, \cite{NC-stateleg-report}.

The key idea, eloquently expressed by Jordan Ellenberg in \cite{Ellenberg19}, is that the opposite of ``gerrymandering" is simply ``not gerrymandering," i.e., drawing maps in accordance with all relevant legal criteria, but incorporating no other political bias into the process.  Depending on the political geography of a state or other region, a map drawn by ``not gerrymandering" may or may not result a map that satisfies any particular standard regarding proportional representation, partisan symmetry, efficiency gap, or other measures.  

As one extreme example, Moon Duchin, et.~al.~showed in \cite{MGGGMass} that over a wide range of historical voting data, {\bf every possible} Congressional district map for Massachusetts would produce a 9-0 Democratic Congressional delegation, even for voting patterns where Republicans receive between 30\% and 40\% of the vote in the state.  This highly non-proportional outcome is not due to partisan gerrymandering, but rather to the fact that Republican voters are too evenly spread out throughout the state, making it impossible to draw even one district where they form the majority.

So, how can we understand what ``not gerrymandering" looks like?  Chen and Rodden's idea was to use a random process to construct a sample, now often called an ``ensemble," of districting plans.  Each plan in the ensemble is built by randomly grouping together small geographic units to form contiguous districts, in such a way that resulting plan is ``valid" under whatever constraints are imposed by legal requirements.  (Typical requirements include that districts be contiguous, have close-to-equal population, be reasonably compact, satisfy requirements of the Voting Rights Act, keep communities of interest together within districts, etc.)  For analysis of partisan gerrymandering, the small geographic units used as building blocks for districts are typically voting precincts, since these are the smallest units for which election results are reported.  

Once the ensemble of plans has been constructed, actual population and voting data from past elections is used to compute measures of interest for each plan in the ensemble; for instance, with this particular plan and this voting data, how many Congressional seats would each party have won? This process produces a statistical range of outcomes for each measure, to which any proposed plan may be compared.  If a proposed plan is an extreme outlier compared to the ensemble, this may indicate that the plan was drawn with some specific goal in mind beyond the stated legal criteria for valid plans, such as partisan gerrymandering.  

This approach can be used not only to flag possible partisan gerrymanders, but also to identify cases where a non-proportional outcome is in fact {\bf not} the result of gerrymandering.
In the Massachusetts example mentioned above, measures such as the efficiency gap would flag {\bf any} district plan as a partisan gerrymander, but ensemble analysis would show that a 9-0 Democratic delegation is, in fact, the expected outcome.

This idea has been gaining traction in redistricting litigation in the last few years.  For instance, Jonathan Mattingly, et. al. performed detailed ensemble analyses of North Carolina's Congressional \cite{NC-cong-report} and state \cite{NC-stateleg-report} legislative district plans that played key roles in the court cases \cite{Rucho} and \cite{CommonCause}, and Moon Duchin's ensemble analysis \cite{PA-report} of Pennsylvania's Congressional Districts played a similar role in \cite{LWV}.  Similar work can be found in Wesley Pegden's expert reports for Pennsylvania \cite{Pegden-PA} and North Carolina \cite{Pegden-NC}.

So, how do we build an ensemble of districting plans?  The construction of a representative sample of plans is not as simple as it might sound; in most cases, the space of possible districting plans is astronomically large and impossible to describe in detail.  Indeed, it is not even clear what is meant by a ``representative" sample; for instance, should the sample be drawn uniformly from the space of all valid plans?  While this might sound like an obviously desirable goal, it turns out to be more problematic than one might expect. The first obstacle is that, as is shown in \cite{najt2019complexity}, uniform sampling from this space is NP-hard; this means that in practice, it may be impossible (or at least highly impractical).  But more interestingly, it turns out that ``most" legally valid plans have undesirable properties such as long, snaky districts, and are in fact quite different from typical human-drawn plans. So it may actually be preferable to use a sampling method that is more likely to sample ``nice" plans, whatever that may mean.  

In any case, in order to understand the properties of a sample generated by any particular method, it is important to have some understanding of how the method affects the properties of plans in the ensemble.
Some examples of methods that have been used to construct samples of plans include floodfill methods \cite{CDO00}, randomized aggregation of voting precincts into districts \cite{CR13}, and optimization algorithms \cite{LCW16}, but the statistical properties of ensembles constructed by these methods have not been thoroughly explored, and very little is known about the sampling distributions for these methods.

One way to sample from a (generally unknown) probability distribution on a large space is to model the space as a graph, with vertices representing elements of the space (in this case, valid districting plans) and with two vertices connected by an edge if they are considered ``close" in some sense.  Then we can build an ensemble by taking a random walk on the graph and adding each plan that we encounter to the ensemble. This idea leads to a family of methods known as Markov Chain Monte Carlo (MCMC), which have a well-developed statistical theory and a long history of applications (see, e.g., \cite{Diaconis09}).  MCMC methods were first applied in the context of redistricting by Mattingly and Vaughn in \cite{MV14}, and they have rapidly become the predominant approach to ensemble sampling for mathematicians studying redistricting.

The most important property of Markov chains that makes them so attractive for ensemble sampling is that a sufficiently long run of an ergodic Markov chain is theoretically guaranteed to produce an ensemble that accurately represents a specific probability distribution on the entire space, regardless of the starting point of the chain.  The main issues that arise are:
\begin{itemize}
\item How well can we describe the probability distribution that the Markov chain is sampling from?  This distribution is determined by the edge structure of the graph (i.e., which plans are adjacent to each plan in the graph) and the algorithm used to decide how to move from one state to an adjacent state, but for large graphs it is generally impractical to describe this distribution explicitly.
\item How long does the Markov chain run need to be in order to guarantee an approximately representative sample?  This length is related to the ``mixing time" for the chain; there is currently no known way to determine it rigorously, and in practice heuristic methods are used to determine experimentally when a chain appears to be sufficiently long.
\end{itemize}

In \S \ref{flip-recom-sec}, we will describe two of the most common MCMC methods used to construct ensembles of districting plans: the Flip method and the ReCom method.  In \S \ref{newprops-sec}, we will present some quantitative experimental evidence that provides insight into the properties of the probability distribution associated to the ReCom method.

\section{MCMC methods: Flip and ReCom}\label{flip-recom-sec}

\begin{figure}[h]
\includegraphics[width=3in]{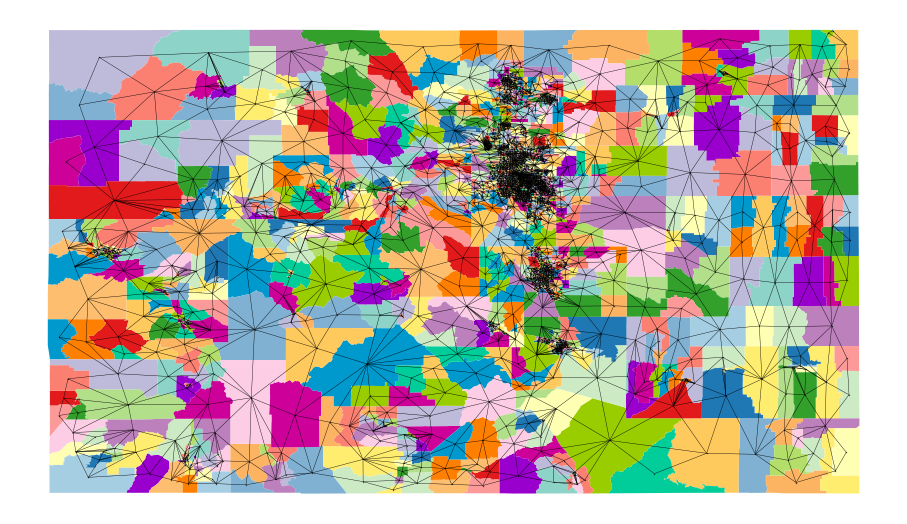}
\includegraphics[width=3in]{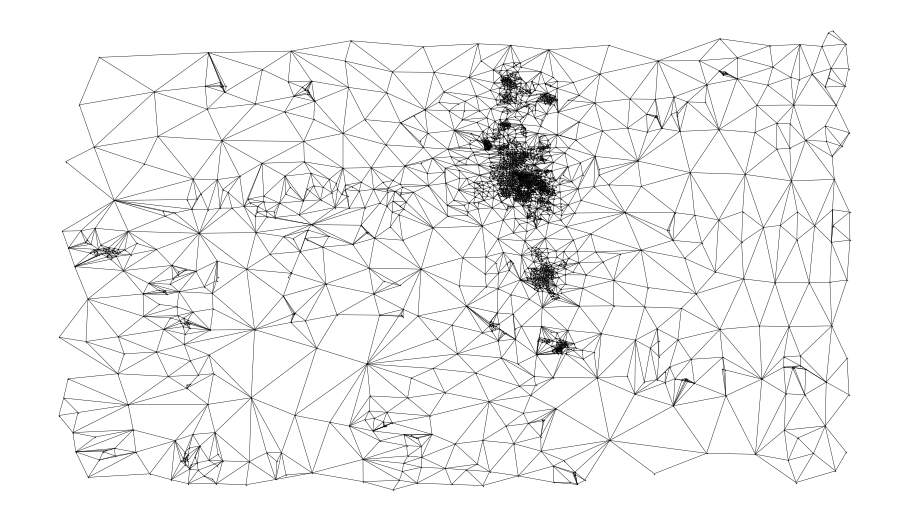}
\caption{Colorado precinct map and dual graph}
\label{CO-dual-graph}
\end{figure}

\begin{figure}[h]
\includegraphics[width=3in]{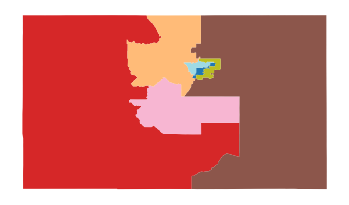}
\includegraphics[width=3in]{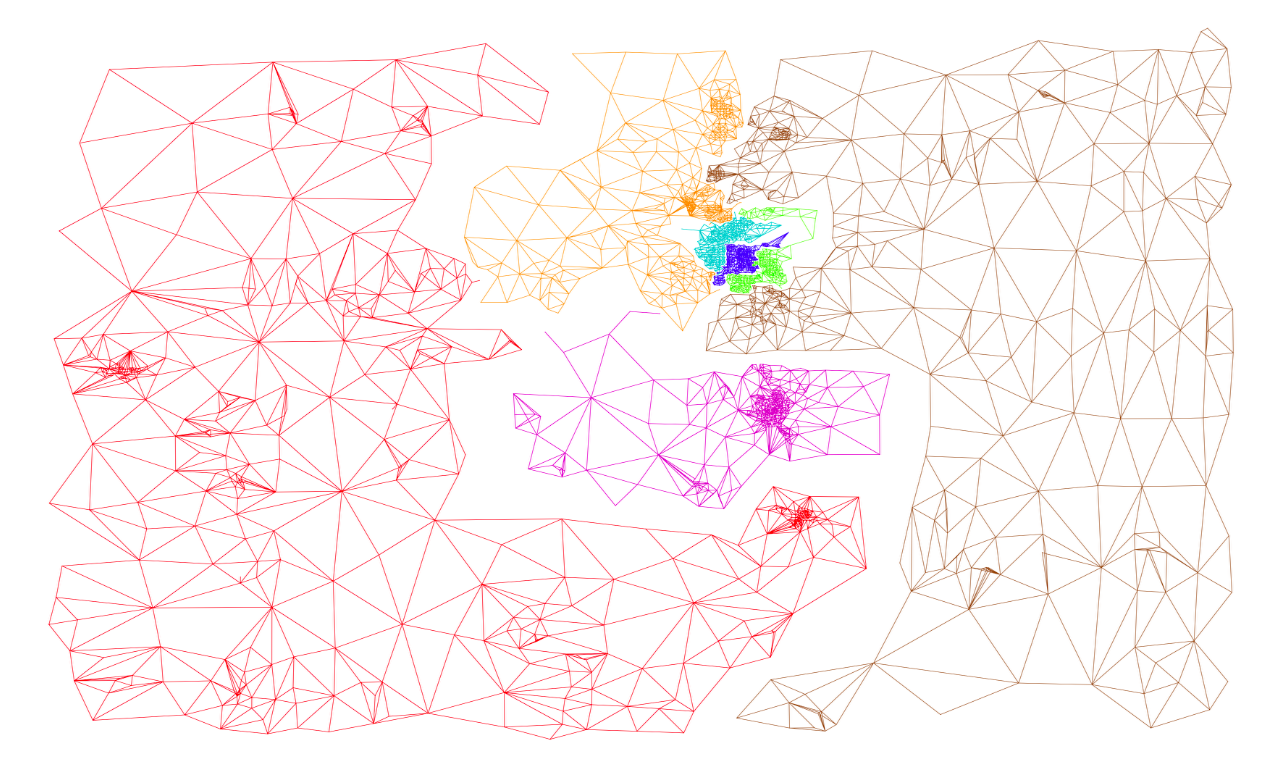}
\caption{Colorado's Congressional districts and corresponding dual subgraphs}
\label{CO-Congressional-dual-graphs}
\end{figure}

In order to start building districting plans, a geographic region (typically a state) is divided into small geographic units, such as voting precincts. This underlying geography is modeled by a graph known as the {\bf dual graph} of the geographic subdivision, where each precinct is represented by a vertex, and two vertices are connected by an edge if the precincts that they represent share a geographic boundary of positive length.\footnote{Note that the dual graph is {\bf not} the graph where the random walk will take place!  The walk takes place on a much larger graph, where each vertex represents a districting plan.  In order to further distinguish between these two graphs, the larger graph where the walk takes place is often referred to as the ``metagraph" of districting plans.}
(See Figure \ref{CO-dual-graph} for an example given by the voting precincts in the state of Colorado.)
A districting plan is then represented by a partition of the dual graph into connected subgraphs, one for each district.  (See Figure \ref{CO-Congressional-dual-graphs}.)

The key ingredient for constructing a Markov chain on the space of districting plans is the choice of a ``proposal function," i.e., a method for moving from one plan to another at each step of the chain.
The simplest method is the one introduced in \cite{MV14}, which has come to be known as the ``Flip" method.  In this method, each step in the Markov chain is performed by randomly selecting a unit on the boundary between two districts and flipping it from one district to the other.  If the resulting plan passes validity checks (e.g., contiguity of districts and appropriate population balance between districts), then the resulting plan is accepted as the next step in the chain.

Mattingly and his collaborators have used variations of this method to construct ensembles for the Wisconsin Assembly \cite{HRM17} and for North Carolina's Congressional and state legislative districts \cite{NC-cong-report}, \cite{NC-stateleg-report}.  The primary advantage of this method is its simplicity; each step in the chain is computationally cheap and very fast to compute.  Moreover, it can be implemented in such a way that the sampling distribution is the uniform distribution on the space of districting plans (see, e.g., \cite{Chikina2860}, \cite{DDS19}).  But it has significant disadvantages as well; as mentioned above, uniform sampling tends to produce plans with very snaky, non-compact districts that are atypical of human-drawn plans, and so the uniform distribution is actually not the most desirable sampling distribution for applications.  Moreover, the theoretical complexity of the uniform sampling problem is borne out in practice; experimental results show that Flip chains exhibit extremely slow mixing (see, e.g., \cite{VA-report}). It should be noted that the Flip algorithm can be embellished with Metropolis-Hastings-style score functions and simulated annealing techniques in order to increase the sampling probability of more desirable plans, and indeed Mattingly and his collaborators have used such techniques extensively in the construction of their ensembles.  The slow mixing time, however, remains a significant issue.

\begin{figure}[h]
\begin{center}
\includegraphics[width=5in]{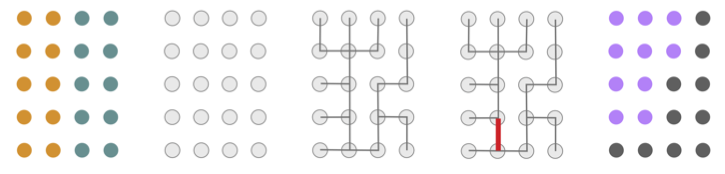}
\end{center}
\caption{A ReCom step (Figure 4 in \cite{DDS19}; used with permission.)}
\label{recom-step-fig}
\end{figure}

In \cite{VA-report}, a new proposal function was introduced and dubbed ``ReCom," motivated by the biological metaphor of recombinant DNA; Markov chains constructed by this method are explored more fully in \cite{DDS19}.
At each step of this Markov Chain, the ReCom algorithm randomly selects a pair of adjacent districts (with probability proportional to the length of the boundary between the districts) and merges the two subgraphs corresponding to these districts into a single graph.  Next, it generates a random spanning tree for the merged graph, chosen uniformly from the set of all spanning trees on the merged graph via Wilson's loop-erased random walk algorithm \cite{Wilson96}. (Alternatively, a minimal spanning tree is generated using Kruskal's algorithm with randomized weights, which produces qualitatively similar results and is computationally faster.)  Finally, it looks for edges that can be cut in order to create two new subgraphs that each satisfy the desired population constraint.  (District contiguity is automatic with this method.)  If no such edges exist, then a new tree is drawn; fortunately, the rejection rate at this step is fairly low in practice. If there are multiple such edges, then one is chosen uniformly at random to cut in order to form two new subgraphs corresponding to new districts, and the resulting districting plan is accepted as the next step in the chain.  This process is illustrated in Figure \ref{recom-step-fig}.  Since each ReCom step makes greater changes to a pair of districts than a Flip step, we might hope that a ReCom chain would exhibit faster mixing time than a Flip chain.

What can we say about the probability distribution on the space of districting plans corresponding to the ReCom algorithm?
Wilson's algorithm samples uniformly from the set of all possible spanning trees for a graph, so at each step of the ReCom algorithm, a spanning tree for the two merged districts is chosen uniformly at random.  (The variation using minimal spanning trees samples from a distribution that is similar, but slightly non-uniform.)  However, many different spanning trees may produce the same partition of the merged districts into two new districts, and some partitions may be produced by more spanning trees than others.  

More specifically, given any particular partition $\mathbf{D} = (D_1, D_2)$ of the merged graph into two 
districts $D_1, D_2$ of equal population, any spanning tree that produces this partition must consist of separate spanning trees for each of $D_1$ and $D_2$, connected by a single edge that joins one vertex of $D_1$ to one vertex of $D_2$.  (This single edge connecting the two smaller trees is the one that will be cut to form the two districts.)  
Therefore, the number of spanning trees that produce this partition is equal to the product 
\begin{equation} \label{num-trees-per-plan}
P_{\mathbf{D}} = T_{D_1} T_{D_2} E_{\mathbf{D}},
\end{equation}
where $T_{D_j}$ denotes the number of spanning trees for the subgraph $D_j$ and $E_{\mathbf{D}}$ denotes the number of edges connecting $D_1$ to $D_2$.
Thus we see that the probability of choosing any particular bipartition $\mathbf{D} = (D_1, D_2)$ is proportional to this product. 

It is less clear exactly how this process of redrawing two districts at a time translates into a specific probability distribution on districting plans with more than two districts, but it still seems intuitively clear that the ReCom method samples from a non-uniform distribution on the space of plans and tends to produce districts whose underlying graphs have large numbers of spanning trees.  It is also fairly easy to see that districts with long, thin tendrils tend to have fewer spanning trees than more ``compact," regularly shaped districts, so we might expect that the ReCom method would tend to produce plans with relatively compact districts.  

Compactness of districts is an important notion, both mathematically and legally, but it is also a vague one. Many different definitions appear in the literature, mostly involving some relationship between a district's perimeter and its area.  There are a variety of issues with these measures, and different measures may disagree as to which districts are more compact than others; see \cite{BNS20}, \cite{DT18} for discussions of some of these issues. 

In \cite{DT18}, Duchin and Tenner discussed a discrete measure for compactness, based on a discrete analog of district perimeter.  Given a dual graph and a partition of the graph into districts, an edge of the graph is called a {\bf cut edge} if its endpoints belong to different districts.  The number of cut edges in a plan is a discrete proxy for the sum of the lengths of the boundaries between districts, and it avoids some of the problems that are inherent is continuous measures; see \cite{DT18} for more details.


The first experimental evidence regarding compactness properties of the ReCom method appeared in \cite{VA-report}, where a comparison of ensembles generated by the Flip and ReCom methods seemed to indicate that:
\begin{itemize}
\item Mixing time appears to be much shorter for ReCom than for Flip.  For the particular scenario analyzed in \cite{VA-report}, ReCom chains starting from a variety of initial seed plans appeared to be well-mixed with regard to summary statistics after 20,000 steps, whereas Flip chains starting from the same seed plans failed to achieve convergence even after 10 million steps.
\item ReCom tends to produce plans with significantly more compact districts (as measured by the total number of cut edges in each plan) than Flip.  
\end{itemize}

The primary goal of this paper is to explore the relationship between spanning trees and cut edges in greater detail, with an eye towards better understanding compactness properties of districts produced by the ReCom method.  We will focus on plans with only two districts, where we know that the ReCom sampling distribution on district plans is proportional to the function $P_{\mathbf{D}} $ in equation \eqref{num-trees-per-plan}.  Our experiments provide strong evidence that for both ideal and real-world dual graphs, this function is related to the number $E_{\mathbf{D}}$ of cut edges by
\begin{equation} \label{PD-formula} 
P_{\mathbf{D}}  \approx C e^{-k E_{\mathbf{D}}} 
\end{equation}
to a good approximation, where $C$ and $k$ are constants depending on the underlying graph.  This relationship has a few important practical implications:
\begin{enumerate}
\item For large districts, the number of spanning trees is computationally intensive to compute, whereas the number of cut edges is simple to compute.  So this relationship may allow approximate relative sampling probabilities for particular plans to be computed more efficiently.
\item The number of cut edges is a much easier quantity to visualize and to explain than the number of spanning trees, so if the probability distribution can be described in these terms, it may be easier to explain to a broad audience.  This is particularly important for political and legal applications of these methods.
\end{enumerate}

\section{Quantifying the relationship between spanning trees and cut edges}\label{newprops-sec}

In this section we will explore the relationship between the function $P_\mathbf{D}$ and the number of cut edges $E_\mathbf{D}$ for bipartitions of three graphs:
\begin{enumerate}
\item a $12 \times 12$ grid graph;
\item a $6 \times 24$ grid graph;
\item the dual graph for voting precincts in Boulder County, CO.
\end{enumerate}
The two grid graphs will be used to compare results for graphs with similar internal structure; both graphs have 144 vertices and all internal vertices have degree 4, but they have different global dimensions and compactness properties.  The Boulder County graph has 234 vertices with degrees ranging from 2 to 24 and average degree $5.47$; it will be used to explore how well the observed relationship holds up for less regular, ``real-world" dual graphs.

In order to compute the function $P_\mathbf{D}$ for a given partition $\mathbf{D} = (D_1, D_2)$, we need to know the number of spanning trees $T_{D_i}$ for each of the districts $D_i$.  Fortunately, a remarkable theorem of Kirchhoff (see, e.g., \cite{MM11}) provides a way to count the number of spanning trees for any finite graph $G$.  Suppose that $G$ has $N$ vertices $(v_1, \ldots, v_N)$.  The {\bf graph Laplacian} $\Lambda_G$ of $G$ is the $N \times N$ matrix whose entries $\Lambda_{ij}$ are defined as follows:
\begin{itemize}
\item If $i=j$, then $\Lambda_{ii} = \deg(v_i)$, the degree of the vertex $v_i$.
\item If $i \neq j$, then $\Lambda_{ij} = -1$ if $G$ contains an edge joining $v_i$ to $v_j$, and $\Lambda_{ij} = 0$ otherwise.
\end{itemize} 
Kirchhoff's theorem says that the number of spanning trees of $G$ is equal to
\begin{equation} \label{Kirchhoff-eq} 
T_G = \det(\Lambda_G'), 
\end{equation}
where $\Lambda_G'$ is any $(N-1) \times (N-1)$ minor of $\Lambda_G$.  Equivalently, 
\[ T_G = \frac{1}{N} \lambda_1 \lambda_2 \cdots \lambda_{N-1}, \]
where $\lambda_1, \ldots, \lambda_{N-1}$ are the nonzero eigenvalues of $\Lambda_G$.

In principle, Kirchhoff's formula makes it possible to compute precisely the number of spanning trees for a given graph.  But in practice, computing determinants for large matrices is computationally intensive and prone to overflow errors.  This is the main reason why we have confined our experiments to relatively small graphs in order to avoid computational difficulties.

\subsection{The $12 \times 12$ grid graph}\label{12x12-sec}

We will start by investigating the case of partitioning a $12 \times 12$ grid graph into two districts with approximately equal numbers of vertices. For purposes of this paper, we will refer to the number of vertices in a district as its ``population."
We allow small deviations from exact population equality because many spanning trees cannot be cut into exactly equal parts, and the computational time required to generate a representative sample is much longer if perfect equality is required.  This also more accurately represents the redistricting problem, where perfect population equality between districts is generally not achievable in practice.  Specifically, we allow the population of each district to differ by up to $2.5\%$ from the ideal population of 72 vertices; this is in keeping with traditional districting criteria for state legislatures, which typically allow for population differences of up to $5\%$ between the largest and smallest districts.

\begin{figure}[h]
\begin{center}
\includegraphics[width=1.7in]{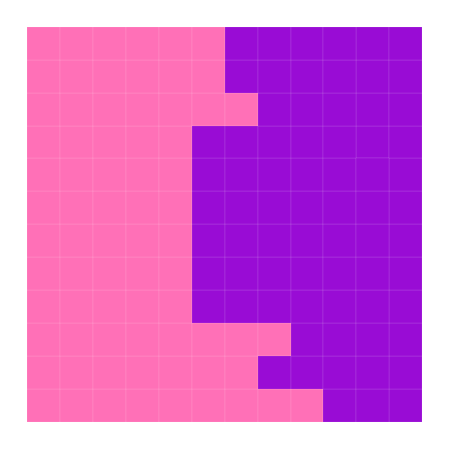}
\includegraphics[width=1.7in]{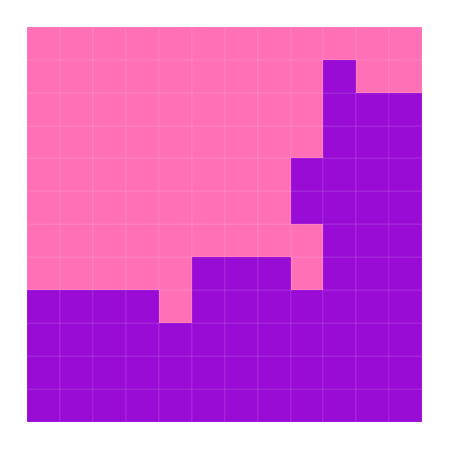}
\includegraphics[width=1.7in]{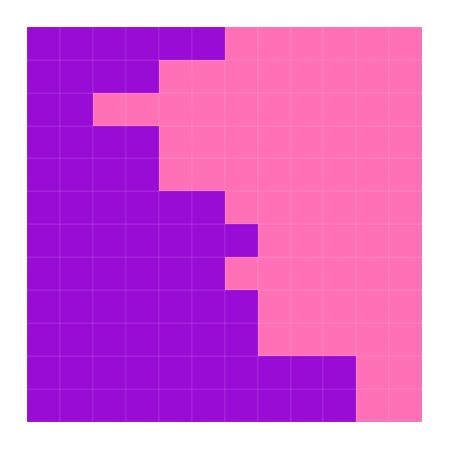}
\end{center}
\caption{Sample plans from ReCom chain on $12 \times 12$ grid graph}
\label{12x12grid-ReCom-sample-plans-fig}
\end{figure}

First, we constructed an ensemble of 100,000 plans via a ReCom chain of 100,000 steps.  With only two districts, each step of this chain is independent of the previous step, so this chain may be regarded as 100,000 independent samples from the probability distribution on the space of plans that is proportional to the function $P_\mathbf{D}$ of equation \eqref{PD-formula}.  Some examples of plans from this chain are shown in Figure \ref{12x12grid-ReCom-sample-plans-fig}.

\begin{figure}[h]
\begin{center}
\includegraphics[width=2in]{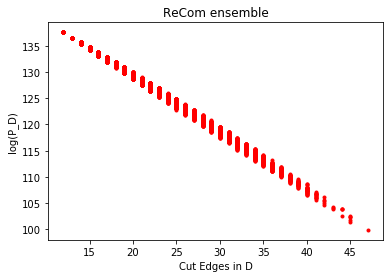}
\includegraphics[width=2in]{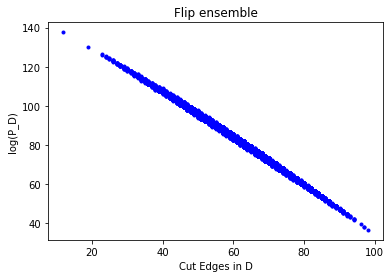}
\includegraphics[width=2in]{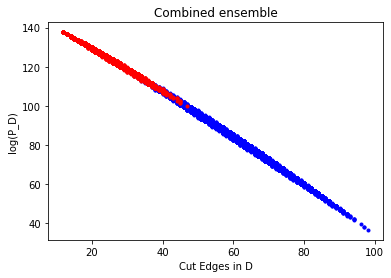}
\end{center}
\caption{$\log(P_\mathbf{D})$ vs.~cut edges for $12 \times 12$ grid graph}
\label{12x12grid-scatterplots-fig}
\end{figure}

The first plot in Figure \ref{12x12grid-scatterplots-fig} is a scatterplot of the function $\log(P_\mathbf{D})$ vs.~the number $E_\mathbf{D}$ of cut edges for all plans in this ensemble.  The graph depicts a clear, approximately linear relationship between these quantities.  More quantitatively, the results of a linear regression analysis on this data are given in the first row of Table \ref{12x12grid-regression-data-table}.  (In order to capture the full range of behavior, this table also includes the maximum error between the observed and predicted values for $\log(P_\mathbf{D})$ over the entire ensemble.)  This data shows that for this graph, the linear relationship between these functions is very strong, and thus that the formula \eqref{PD-formula} describes the probability distribution $P_\mathbf{D}$ associated to the ReCom method in terms of the number of cut edges to a high degree of accuracy.

\begin{figure}[h]
\begin{center}
\includegraphics[width=1.7in]{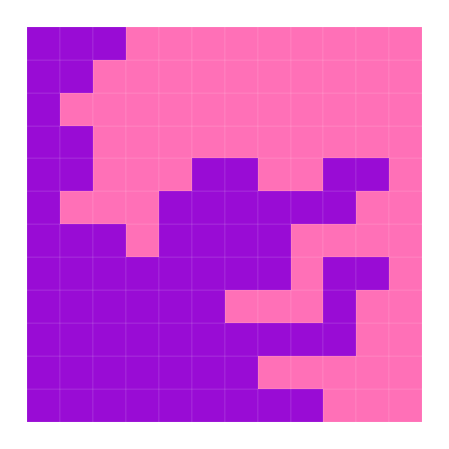}
\includegraphics[width=1.7in]{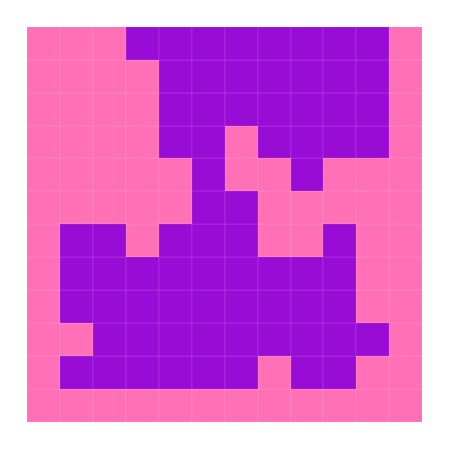}
\includegraphics[width=1.7in]{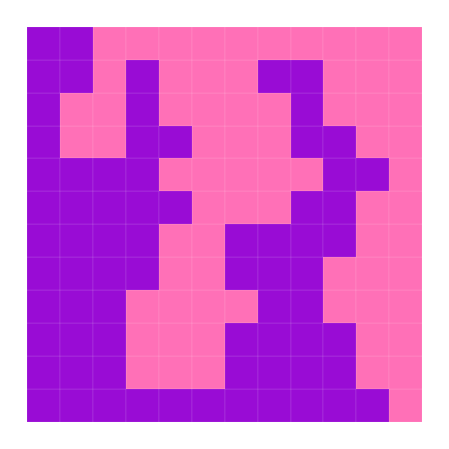}
\end{center}
\caption{Sample plans from Flip chain on $12 \times 12$ grid graph}
\label{12x12grid-Flip-sample-plans-fig}
\end{figure}

Next, we were curious to see whether this relationship was consistent more broadly across the entire space of plans, and not just for the more compact plans that the ReCom method tends to favor.  So we constructed a second ensemble of 100,000 plans via a Flip chain of 1,000,000 steps, adding every 10th plan in the chain to the ensemble.\footnote{Note that we used the basic Flip method, not the modified version that samples uniformly from the space of plans.  We make no claims of uniform sampling here; we simply wanted to create a second ensemble of plans that were less compact than those produced by the ReCom method.}
Some examples of plans from this chain are shown in Figure \ref{12x12grid-Flip-sample-plans-fig}.

\begin{figure}[h]
\begin{center}
\includegraphics[width=1.7in]{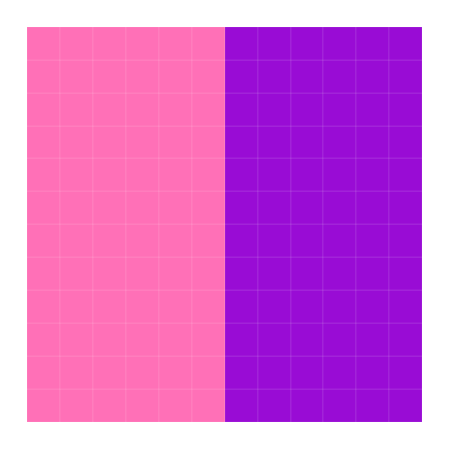}
\includegraphics[width=1.7in]{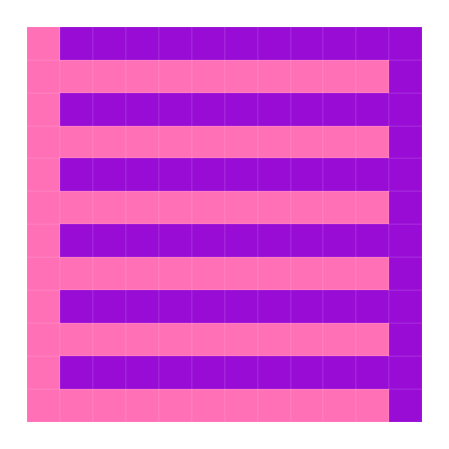}
\end{center}
\caption{Maximally and minimally compact seed plans for Flip chain on $12 \times 12$ grid graph}
\label{12x12grid-Flip-seed-plans-fig}
\end{figure}

\begin{remark}
We actually ran two Flip chains, one starting from a maximally compact seed plan and one from a minimally compact seed plan.  (See Figure \ref{12x12grid-Flip-seed-plans-fig}.)  After a transitory period of only about 100 steps, both chains appeared to settle down and explore the same region of the sample space, and their linear regression statistics were essentially identical.  Here we will report the results for the chain starting from the compact seed.
\end{remark}

The second and third plots in Figure \ref{12x12grid-scatterplots-fig} are scatterplots of the function $\log(P_\mathbf{D})$ vs.~the number $E_\mathbf{D}$ of cut edges for all plans in the Flip ensemble and for all plans in the combined ensemble of 200,000 plans, respectively.  These graphs seem to indicate that the linear relationship between these quantities remains strong over the entire sample space.  The second and third rows of Table \ref{12x12grid-regression-data-table} show the linear regression statistics for the Flip ensemble and the combined ensemble, respectively.  Taken together with the linear regression statistics for the ReCom ensemble, we see that the slopes indicate some slight downward concavity to this function over the entire space; nevertheless, the linear fit remains very strong, even for the combined ensemble.

\begin{table}[h]
	\centering
	\begin{tabular}{| l |c |c | c | c | c | }
		\hline
		Ensemble & Slope & Intercept &  Correlation coefficient  & Mean squared error & Maximum error\\ 
		\hline
		ReCom & -1.06756  &  150.72129  & -0.99835   & 0.06148  &  1.52213\\ 
		\hline
		Flip & -1.19456	 &  155.80851  & -0.99897   &  0.23589 &  3.88778  \\
		\hline
		Combined & -1.14164 & 152.39838   & -0.99975    &  0.31117  &  3.82996 \\
		\hline 
	\end{tabular}
	\vspace{0.1in}
	\caption{Linear regression statistics for ensembles on $12 \times 12$ grid graph}
	\label{12x12grid-regression-data-table}
\end{table}

\begin{figure}[h]
\begin{center}
\includegraphics[width=4in]{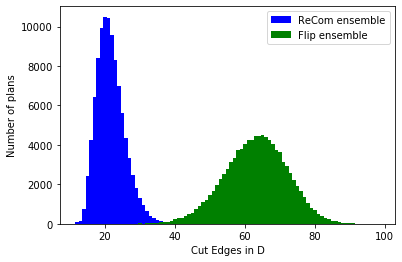}
\end{center}
\caption{Cut edges frequency distributions for $12 \times 12$ grid graph}
\label{12x12grid-cut-edges-hist-fig}
\end{figure}

Finally, we plotted histograms showing the frequency distributions for cut edges for both the ReCom and Flip ensembles; these plots are shown in Figure \ref{12x12grid-cut-edges-hist-fig}.  These histograms provide some additional insight into the structure of the space of plans: Even though the individual plans with the fewest cut edges have the highest ReCom sampling probability, there are very few such plans in the sample space, and so their overall occurrence in the ensemble is very low.  By contrast, there are many more plans with slightly more cut edges---indeed, sufficiently more that such plans are much more likely to be sampled, even though the individual sampling probability for any such plan is much smaller.  The ReCom histogram in Figure \ref{12x12grid-cut-edges-hist-fig} gives some indication of the balance between these competing factors across the entire space of plans; the peak of the histogram occurs for plans with 20-21 cut edges, slightly less than twice the minimum value of 12. Meanwhile, the Flip histogram in Figure \ref{12x12grid-cut-edges-hist-fig} confirms that the Flip method samples primarily from much less compact plans than the ReCom method.

\subsection{The $6 \times 24$ grid graph}\label{6x24-sec}

The next graph that we chose to investigate is a $6 \times 24$ grid graph.  This graph has the same number of vertices as the $12 \times 12$ grid graph of section \ref{12x12-sec}, with very similar internal structure; the main difference between these two graphs is their global shape and relative compactness.  This difference can be described quantitatively in terms of spanning trees; the $12 \times 12$ grid graph has approximately $e^{146.146}$ spanning trees, while the $6 \times 24$ grid graph has approximately $e^{140.419}$ spanning trees---a factor of about $307$ times fewer.

\begin{figure}[h]
\begin{center}
\includegraphics[width=3.2in]{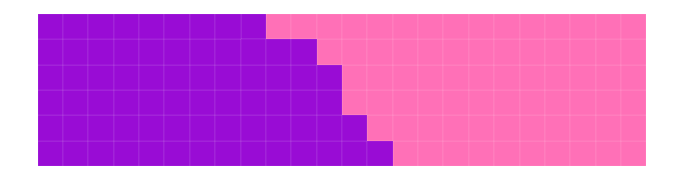}
\includegraphics[width=3.2in]{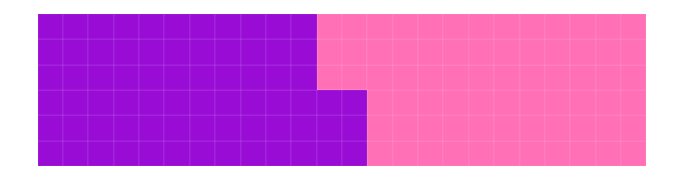}
\includegraphics[width=3.2in]{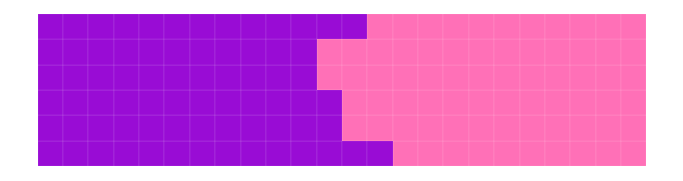}
\end{center}
\caption{Sample plans from ReCom chain on $6 \times 24$ grid graph}
\label{6x24grid-ReCom-sample-plans-fig}
\end{figure}

\begin{figure}[h]
\begin{center}
\includegraphics[width=3.2in]{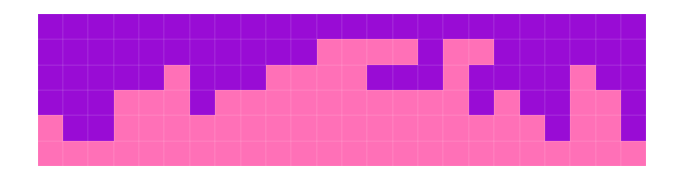}
\includegraphics[width=3.2in]{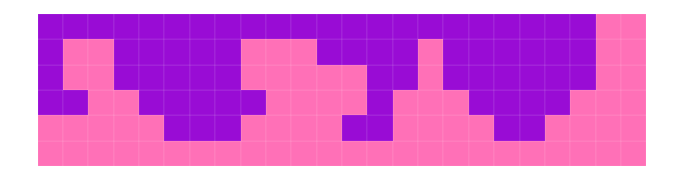}
\includegraphics[width=3.2in]{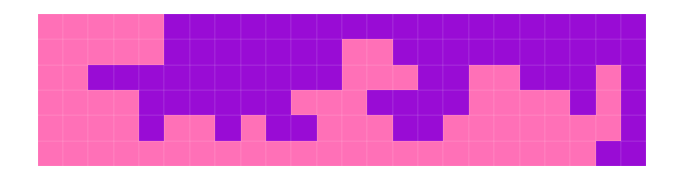}
\end{center}
\caption{Sample plans from Flip chain on $6 \times 24$ grid graph}
\label{6x24grid-Flip-sample-plans-fig}
\end{figure}

We repeated the same analysis as for the $12 \times 12$ grid graph, using ReCom and Flip chains to construct ensembles of 100,000 plans each.  Some examples of plans from the ReCom and Flip ensembles for the $6 \times 24$ grid graph are shown in Figures \ref{6x24grid-ReCom-sample-plans-fig} and \ref{6x24grid-Flip-sample-plans-fig}, respectively.

\begin{figure}[h]
\begin{center}
\includegraphics[width=2in]{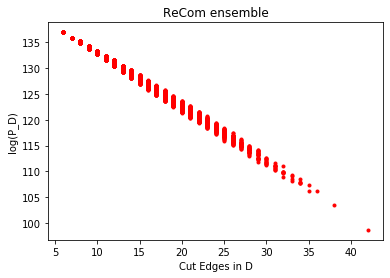}
\includegraphics[width=2in]{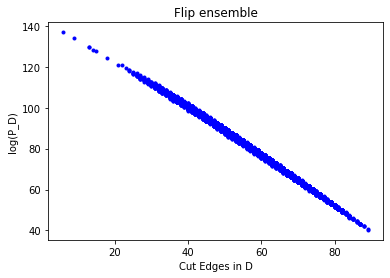}
\includegraphics[width=2in]{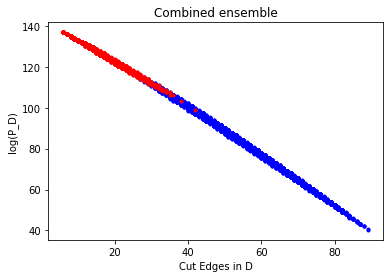}
\end{center}
\caption{$\log(P_\mathbf{D})$ vs.~cut edges for $6 \times 24$ grid graph}
\label{6x24grid-scatterplots-fig}
\end{figure}

Scatterplots of the function $\log(P_\mathbf{D})$ vs.~the number $E_\mathbf{D}$ of cut edges for all plans in the ReCom, Flip, and combined ensembles are shown in Figure \ref{6x24grid-scatterplots-fig}.  Linear regression statistics for all three ensembles are shown in Table \ref{6x24grid-regression-data-table}.  This data is strikingly similar to what we saw for the $12\times 12$ grid graph; the main difference is that the intercepts for each ensemble range from about $6$ to $8$ less than those of the corresponding ensemble for the $12\times 12$ grid graph, indicating that the value of $P_\mathbf{D}$ for these plans is typically about $e^6 - e^8$  (roughly 400 -- 3,000) times less than $P_\mathbf{D}$ for similar plans for the $12\times 12$ grid graph.   These ratios are reasonably similar to those between the total numbers of spanning trees for the two grid graphs, and as such they are exactly what we might expect to see in this case.

\begin{table}[h]
	\centering
	\begin{tabular}{| l |c |c | c | c | c | }
		\hline
		Ensemble & Slope & Intercept &  Correlation coefficient  & Mean squared error & Maximum error\\ 
		\hline
		ReCom & -1.02535  &  143.14522  &  -0.99805  & 0.04329  &  1.61562  \\ 
		\hline
		Flip & -1.20926	 &  149.48318  & -0.99910   & 0.18060  &  5.33475  \\
		\hline
		Combined & -1.11894  & 144.26036   &  -0.99968   & 0.44947   & 4.40286  \\
		\hline 
	\end{tabular}
	\vspace{0.1in}
	\caption{Linear regression statistics for ensembles on $6 \times 24$ grid graph}
	\label{6x24grid-regression-data-table}
\end{table}

\begin{figure}[h]
\begin{center}
\includegraphics[width=4in]{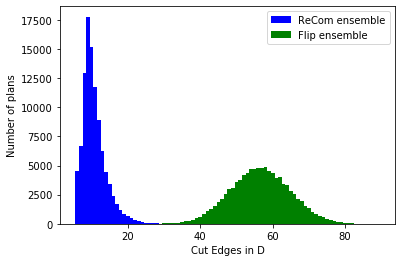}
\end{center}
\caption{Cut edges frequency distributions for $6 \times 24$ grid graph}
\label{6x24grid-cut-edges-hist-fig}
\end{figure}

The frequency distributions for cut edges shown in Figure \ref{6x24grid-cut-edges-hist-fig}, however, tell a slightly different story.  The ReCom ensemble for the $6 \times 24$ grid graph has a stronger bias towards plans with fewer cut edges compared to what we saw for the $12 \times 12$ grid graph; the peak of the histogram occurs for plans with 9 cut edges, only $1.5$ times the minimum value of 6, which is itself half the minimum value of 12 for the $12 \times 12$ grid graph.  Meanwhile, the Flip histogram for the $6 \times 24$ grid graph is qualitatively very similar to that for the $12 \times 12$ grid graph;  the slightly smaller value for the number of cut edges at the peak of the histogram is explained by the fact that the $6 \times 24$ grid graph has slightly fewer internal edges than the $12 \times 12$ grid graph (264 vs.~258).  This suggests that compactness properties of plans produced by the Flip method may be less affected by the global shape of the graph than those produced by the ReCom method.

Taken together, we see that the gap between the peaks of the ReCom and Flip histograms for the $6 \times 24$ grid graph is substantially wider than that for the corresponding $12 \times 12$ grid graph histograms.
This indicates that there are proportionally more plans with small numbers of cut edges for the $6 \times 24$ grid graph than for the $12 \times 12$ grid graph, so that the higher sampling probability for such plans is less attenuated by the increasing number of plans with more cut edges in the overall frequency distribution.  Intuitively, this can be visualized by comparing the sample ReCom plans in Figure \ref{6x24grid-ReCom-sample-plans-fig} with those in Figure \ref{12x12grid-ReCom-sample-plans-fig}: The less compact structure of the overall graph gives ReCom fewer choices for the shapes of districts with many spanning trees, leading it to prefer plans that divide the graph approximately vertically and have only slightly more than the minimum number of cut edges.

\subsection{The Boulder County precinct graph}\label{Boulder-sec}

\begin{figure}[h]
\includegraphics[width=3in]{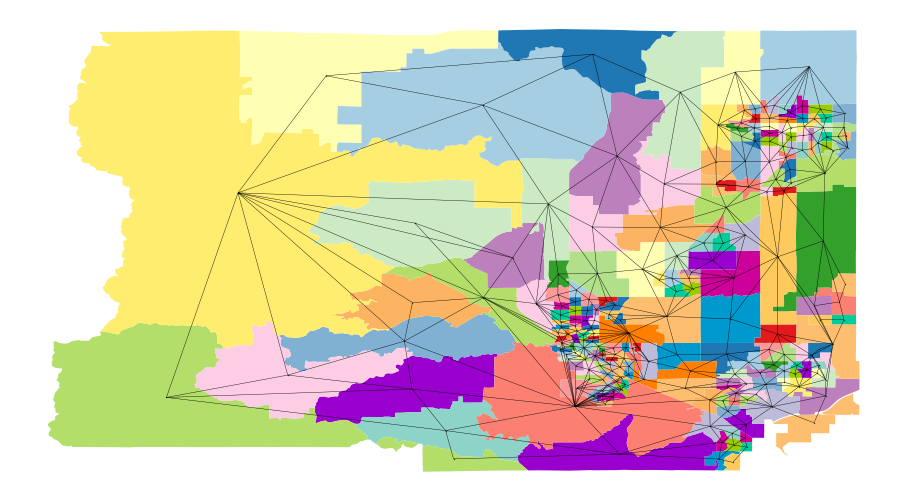}
\includegraphics[width=2.5in]{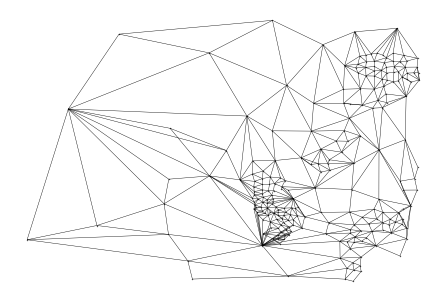}
\caption{Boulder County precinct map and dual graph}
\label{BOCO-dual-graph}
\end{figure}

Finally, we wanted to see whether this relationship would remain intact for a less regular, ``real-world" dual graph.  Somewhat arbitrarily, we chose to examine the precinct graph for Boulder County, CO, where our university is located.  This precinct map and its dual graph are shown in Figure \ref{BOCO-dual-graph}. The graph has 234 vertices, which is larger than the grid graphs of the previous examples but still small enough for the spanning tree computations from Kirchhoff's theorem to remain tractable.  The degrees of the vertices range from 2 to 24, with a mean of $5.47$.

\begin{figure}[h]
\begin{center}
\includegraphics[width=2.8in]{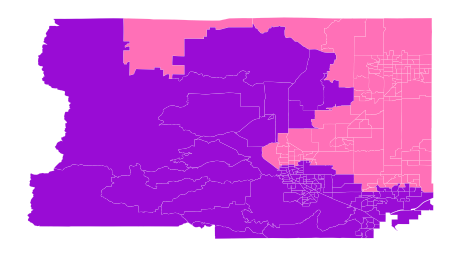}
\includegraphics[width=2.8in]{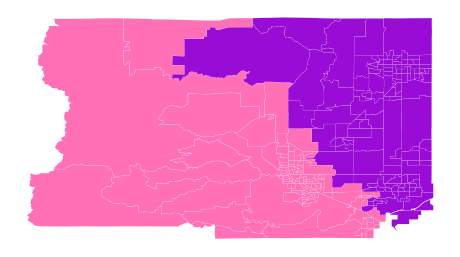}
\includegraphics[width=2.8in]{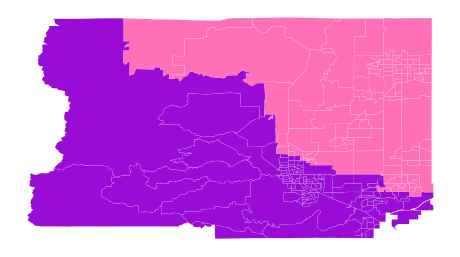}
\end{center}
\caption{Sample plans from ReCom chain on Boulder County precinct graph}
\label{BOCO-ReCom-sample-plans-fig}
\end{figure}

\begin{figure}[h]
\begin{center}
\includegraphics[width=2.8in]{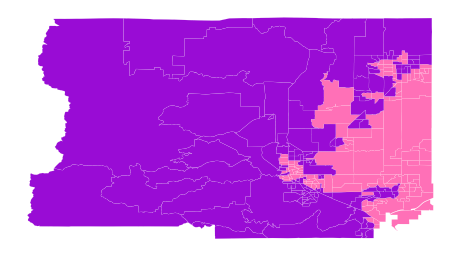}
\includegraphics[width=2.8in]{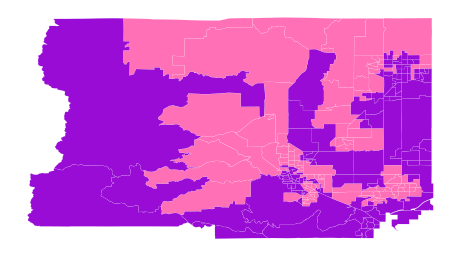}
\includegraphics[width=2.8in]{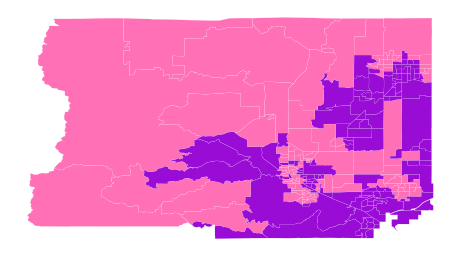}
\end{center}
\caption{Sample plans from Flip chain on Boulder County precinct graph}
\label{BOCO-Flip-sample-plans-fig}
\end{figure}

Again, we used a 100,000 step ReCom chain to construct one ensemble of 100,000 plans, and a 1,000,000 step Flip chain, keeping every 10th step, to construct another ensemble of 100,000 plans.  Some examples of plans from both ensembles are shown in Figures \ref{BOCO-ReCom-sample-plans-fig} and \ref{BOCO-Flip-sample-plans-fig}.

\begin{figure}[h]
\begin{center}
\includegraphics[width=2in]{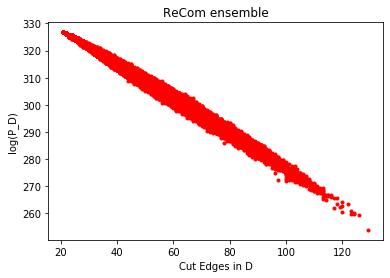}
\includegraphics[width=2in]{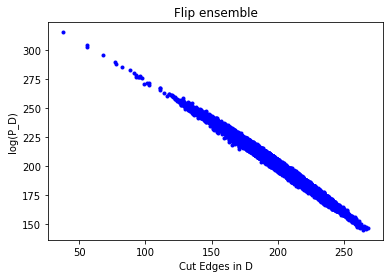}
\includegraphics[width=2in]{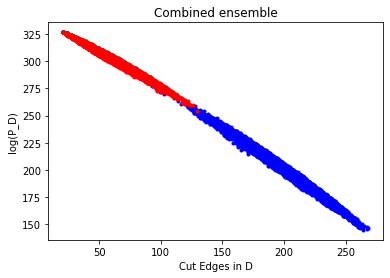}
\end{center}
\caption{$\log(P_\mathbf{D})$ vs.~cut edges for Boulder County precinct graph}
\label{BOCO-scatterplots-fig}
\end{figure}

Scatterplots of the function $\log(P_\mathbf{D})$ vs.~the number $E_\mathbf{D}$ of cut edges for all plans in the ReCom, Flip, and combined ensembles are shown in Figure \ref{BOCO-scatterplots-fig}. Visually, it appears that the same basic linear relationship still holds, but with significantly larger margins of error.
Linear regression statistics, shown in Table \ref{BOCO-regression-data-table}, tell a slightly more nuanced story: The scatterplots depict the full range of values, as shown in the ``Maximum error" columns of Tables \ref{12x12grid-regression-data-table}, \ref{6x24grid-regression-data-table}, and \ref{BOCO-regression-data-table}.  The maximum errors for the Boulder County precinct graph range from approximately 23 to 46 times the maximum errors for the corresponding ensembles for the two grid graphs. The mean square errors for the Boulder County precinct graph, on the other hand, range from approximately 8 to 18 times those for the corresponding ensembles for the two grid graphs; therefore, the mean absolute errors for the Boulder County precinct graph are only on the order of 2.8 to 4.3 times those for the corresponding ensembles for the two grid graphs.  

Additionally, the correlation coefficients for the Boulder County precinct graph ensembles are all  $-0.99444$ or less (compared to $-0.99805$ or less for the grid graphs), so that the coefficients of determination (obtained by squaring the correlation coefficients) are all $0.9889$ or greater (compared to $0.9961$ or greater for the grid graphs).
All of this indicates that the linear relationship between $\log(P_\mathbf{D})$ and $E_\mathbf{D}$ for the Boulder County precinct graph, while slightly weaker than that for grid graphs, is still much stronger than a cursory look at the scatterplots in Figure \ref{BOCO-scatterplots-fig} might suggest.

\begin{table}[h]
	\centering
	\begin{tabular}{| l |c |c | c | c | c | }
		\hline
		Ensemble & Slope & Intercept &  Correlation coefficient  & Mean squared error & Maximum error\\ 
		\hline
		ReCom & -0.62675  &  339.99852  &  -0.99530  & 0.76381  &  6.97904  \\ 
		\hline
		Flip & 	-0.79110 &  359.06389  & -0.99444  &  2.56661 & 13.34141  \\
		\hline
		Combined & -0.71752  & 344.34357   & -0.99944    &  3.50526  & 10.10422 \\
		\hline 
	\end{tabular}
	\vspace{0.1in}
	\caption{Linear regression statistics for ensembles on Boulder County precinct graph}
	\label{BOCO-regression-data-table}
\end{table}

\begin{figure}[h]
\begin{center}
\includegraphics[width=4in]{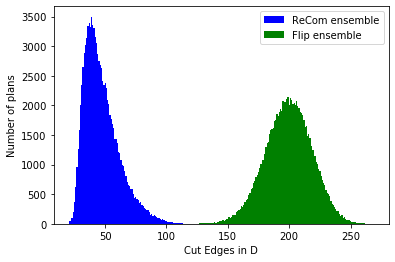}
\end{center}
\caption{Cut edges frequency distributions for Boulder County precinct graph}
\label{BOCO-cut-edges-hist-fig}
\end{figure}

The frequency distributions for cut edges for both ensembles are shown in Figure \ref{BOCO-cut-edges-hist-fig}.  These seem to follow a pattern similar to those for the grid graphs, with the peak of the ReCom histogram occurring at roughly twice the minimum number of cut edges and the Flip histogram appearing qualitatively similar to those for the grid graphs.  Once again, we see that that ``typical" plans produced by the Flip chain tend to be much less compact than those produced by the ReCom chain.

\section{Discussion}\label{Discussion-sec}

While more experimentation on a wider variety of graphs is certainly needed before drawing any firm conclusions, we are very encouraged by the robustness of our results when comparing the Boulder County precinct graph to the two grid graphs.  These experiments seem to provide strong evidence that the probability distribution on bipartitions associated to each individual ReCom step can be described to a good approximation in terms of the number of cut edges in each plan via the formula \eqref{PD-formula}.  

More work needs to be done in order to understand how the constants $C$ and $k$ in equation \eqref{PD-formula} are determined by the properties of the underlying graph. This is especially important for applications to plans with more districts, where each step of the ReCom chain is performed on a different subgraph and so these constants may vary substantially between steps.  The degree to which they vary will have important implications for how effectively the relationship \eqref{PD-formula}
might be applied to describe the sampling probability distribution on plans generated by the ReCom method in terms of cut edges.

There are also other algorithmic choices that may affect the ReCom probability distribution for plans with more than two districts.  Perhaps the most obvious is the algorithm for selecting of a pair of districts to merge at each step.  The original implementation of ReCom chooses a pair of districts with probability proportional to the number of cut edges between them, so it favors the selection of district pairs that share a long geographical boundary. A natural alternative choice would be to select district pairs uniformly at random; how might this distinction impact the resulting probability distribution on the space of plans, particularly with regard to its compactness properties?

Finally, we would like to mention that the MGGG Redistricting Lab has very recently developed a {\bf reversible} version of ReCom \cite{revrecom}, designed so that the sampling probability for a districting plan $\mathbf{D} = (D_1, \ldots D_k)$ is {\bf exactly} proportional to the product $\text{sp}(\mathbf{D})$ of the numbers of spanning trees for each district:
\begin{equation} \label{PD-reversible-formula}
\text{sp}(\mathbf{D}) = T_1 \cdots T_k. 
\end{equation}
This is an important and exciting development, and naturally it comes with a price; specifically, the Markov chain includes an acceptance function that has a very small acceptance probability.  This means that most steps in the chain will be stationary, and an ensemble produced by this chain will contain a large percentage of repeated plans.  These repetitions cannot be removed from the ensemble; they are necessary in order for the probability distribution \eqref{PD-reversible-formula} to remain accurate.  Consequently, mixing times for the reversible version are generally quite a bit longer than those for the original version, and therefore much longer chains and larger ensembles will be required in order to obtain representative samples.  This tradeoff may very well turn out to be well worth the added cost; it remains to be seen exactly how much larger ensembles generated by this method will need to be in practice.   

For our purposes, the availability of the reversible ReCom method presents an exciting opportunity for future work on extending our experiments to plans with more than two districts.  If the relationship \eqref{PD-formula} (or some analog of it) remains valid to a high degree of accuracy for plans with more districts, this would allow us to describe the reversible ReCom sampling probability for any plan in terms of its compactness properties.  As mentioned earlier, such a description would be desirable not only for mathematical reasons, but also for its relative simplicity for purposes of communication with broader audiences.

Even with the addition of reversible ReCom to the library of MCMC methods, we remain convinced of the importance of better understanding the sampling probability distribution associated to the original version of ReCom---particularly as it relates to compactness properties---as ``regular" ReCom ensembles can be generated at lower cost, both in terms of time and computational resources, than those for the reversible version.  Moreover, variations on the original version have already been widely adopted; as one important example, a widely used commercial redistricting software product from Maptitude is using a proprietary version of ReCom in its software for the 2021 redistricting cycle \cite{MaptituteTweet}. 
In future work, we hope to take advantage of reversible ReCom by comparing properties of ensembles generated by both versions, using the known probability distribution for reversible ReCom to better understand similarities and differences between the original and reversible versions.

\bibliographystyle{amsplain}
\bibliography{gerry-bib}

\end{document}